\documentclass[12pt]{article}
\usepackage{epsf}

\newcommand{\bee}{\begin{equation}}
\newcommand{\ee}{\end{equation}}
\newcommand{\bea}{\begin{eqnarray}}
\newcommand{\eea}{\end{eqnarray}}
\newcommand{\R}{\rm I\kern-.2emR}
\newcommand{\C}{\rm \kern.25em\vrule height1.4ex
depth-.12ex width.06em\kern-.31em C}
\newcommand{\N}{{\rm I\kern-.16em N}}
\newcommand{\Z}{{\rm Z\kern-.35em Z}}
\begin{document}                                                                
\begin{flushright}
AZPH-TH-2000-02\\
MPI-PhT/2000-32\\
\end{flushright}
\bigskip\bigskip\begin{center}
{\Huge Discrete Symmetry Enhancement}\\
\vglue 2mm
{\Huge in Nonabelian Models and the}\\
\vglue 3mm
{\Huge Existence of Asymptotic Freedom}
\end{center}
\centerline{Adrian Patrascioiu}
\centerline{\it Physics Department, University of Arizona,}
\centerline{\it Tucson, AZ 85721, U.S.A.}
\centerline{\it e-mail: patrasci@physics.arizona.edu}
\vskip5mm
\centerline{and}
\centerline{Erhard Seiler}
\centerline{\it Max-Planck-Institut f\"ur Physik}
\centerline{\it (Werner-Heisenberg-Institut)}
\centerline{\it F\"ohringer Ring 6, 80805 Munich, Germany}
\centerline{\it e-mail:ehs@mppmu.mpg.de}
\bigskip \nopagebreak

\begin{abstract}
We study the universality between a discrete spin model with
icosahedral symmetry and the $O(3)$ model in two dimensions.
For this purpose we study numerically the renormalized two-point
functions of the spin field and the four point coupling constant.
We find that those quantities seem to have the same continuum
limits in the two models. This has far reaching consequences, because
the icosahedron model is {\it not} asymptotically free in the sense
that the coupling constant proposed by L\"uscher, Weisz and Wolff
\cite{LWW} does not approach zero in the short distance limit.
By universality this then also applies to the $O(3)$ model,
contrary to the predictions of perturbation theory.
\end{abstract}
%\pacs{11.25.Bt, 11.15.Ha, 75.10.Hk}
%1111111111111111111111111111111111111111111111111111111111111111111111
%\narrowtext
\vskip2mm
\section{Introduction}
The subject of the enhancement of a discrete symmetry to a continuous 
one at large distances is important both theoretically and 
phenomenologically.
Indeed, even if there are good grounds to expect that a certain 
material is well described by some model enjoying $O(N)$ symmetry, one 
may wonder what might be the effect of anisotropies. This question was
addressed in 1977 by Jos\'e et al \cite{jose} for the $O(2)$ nonlinear
$\sigma$ model. Nonrigorous renormalization group arguments led them to 
the conclusion that in two dimensions ($2D$) the discrete symmetry 
$Z_N$ should be enhanced to full $O(2)$ invariance if $N\geq 5$ for
$\beta$ (inverse temperature) not too large. For $N$ sufficiently 
large, the occurrence of this phenomenon was proven rigorously by 
Fr\"ohlich and Spencer in 1981 \cite{fs}, who showed that there exists 
a range of temperatures in which spin correlation functions decay 
algebraically and are $O(2)$ invariant. Fr\"ohlich and Spencer proved 
also that a similar phenomenon of discrete abelian symetry enhancement 
occurs in $4D$ gauge theories.

For a long time, the consensus was that no symmetry enhancement should
occur in nonabelian models. The main reason appears to have been the
belief that for continous symmetries, these models exhibit asymptotic
freedom (AF). The discrete models, known rigorously to undergo phase
transitions at nonzero temperature, did not seem likely to be AF, hence 
had to be different. A proposal for nonabelian symmetry enhancement 
came however from Newman and Schulman \cite{ns}. Their argument was 
based on the fact that if the discrete symmetry group is sufficiently 
large, any 4-th  order polynomial invariant under the discrete group is 
also invariant under the continuous group in which it is contained.
While this is an undisputable mathematical fact, the question was why 
4-th order? Their heuristic answer was that the renormalization group 
flow was expected to be free of bifurcations as the dimension $D$ 
was varied between
2 and 4. So if one started just below $D=4$, where the most one could 
have is a $\phi^4$ interaction (higher powers being irrelevant), this
symmetry enhancement should persist down to $D=2$. Since however
in $D=2$ all polynomials in $\phi$ are relevant and it is easy to
write down such polynomials possessing only a discrete symmetry,
and to construct the corresponding $P(\phi)_2$ models, the validity
of the argument of Newman and Schulman remains unclear.

A different heuristic argument in favor of symmetry enhancement for
abelian as well as nonabelian groups was put forward by
Patrascioiu in 1985 \cite{pat}. For spin models, his argument went as 
follows: at sufficiently low temperatures, there exists a phase with long
range order (l.r.o.) because, as Peierls showed long ago, given an 
ordered state, there is not enough free energy to create a domain in 
which the spin points elsewhere. Now consider a model like $Z_5$. As
one increases the temperature, clearly the first abundant
domains to form would be those in which the spin pointed in a direction
immediately neighboring the one chosen by the boundary conditions (b.c.)
for the ordered state. For temperatures not too high, the system
could form domains inside domains of neighboring spin values. This would
be different from the phase at high temperature, where no such 
restriction between adjacent domains would be required. This scenario
does not seem to have anything it do with the model being abelian or not and
Patrascioiu suggested that, since it was known to happen in abelian
models, it must happen also in nonabelian cases.

Except for the papers quoted above, in the 80s, while it was quite
fashionable to replace continous groups with discrete ones in
Monte Carlo simulations, everybody seemed to be convinced that the discrete
and continuous models belonged to different universality classes. Our
interest in the subject was rekindled in 1990 when, together with
Richard, we derived a rigorous inequality relating correlation
functions in the dodecahedron model to those of $Z_{10}$ \cite{prs}. Our result
was that for any $\beta$ the dodecahedron model is more ordered than 
$Z_{10}$ at $0.607^2\beta$. Since it is was pretty well established
that $Z_{10}$ possess an extended intermediate phase which is $O(2)$
invariant, our inequality implied that provided 
$\beta_m({\it D})>\beta_c(Z_{10})$ the dodecahedron must also
possess an intermediate massless phase. Here $\beta_m({\it D})$ denotes
the onset of the l.r.o. phase in the dodecahedron and $\beta_c(Z_{10})$
the onset of algebraic decay in $Z_{10}$. We determined numerically these
values \cite{dod} and concluded that the dodecahedron seemed to
possess an intermediate massless phase for $2.15<\beta<2.8$. We 
conjectured that this phase must enjoy full $O(3)$ invariance.
 
Intrigued by our findings for the dodecahedron, in the early 90s
we looked numerically at the other regular polyhedra. While the cube is
obviously equivalent to 3 uncoupled Ising models, hence not a good candidate
for exhibiting $O(3)$ invariance, the other 3 regular polyhedra (platonic
solids) a priori are.  Actually in the
scenario advocated by Patrascioiu in 1985 \cite{pat}, the tetrahedron
should not be able to simulate spin waves since its spin gradient can
take only one nontrivial value (in fact it is nothing but the 4 state
Potts model). The octahedron and the icosahedron could. Our numerics
suggested that the octahedron had a first order transition. For
the icosahedron the MC data suggested a second order
transition from the high temperature phase to the low temperature
phase exihibiting l.r.o.; in particular there did not seem to be an
extended massless phase, as for the dodecahedron.

That the discrete icosahedral symmetry may be enhanced to $O(3)$
began to become manifest in 1998 when we began extensive numerical 
investigations of the contiuum limit of the spin 2-point function 
versus $p/m$ \cite{pl1} in the dodecahedron model.
The original motivation of that study was to compare the 
lattice continuum limit with the form factor (FF) prediction of Balog
and Niedermaier \cite{bn}. Since it was not to be expected that
the latter could possibly describe the continuum limit of the
dodecahedron model, we decided to take data on this model too.
To our surprise, the $O(3)$ data seemed to agree with both the 
FF prediction and with the dodecahedron. Since Balog and Niedermaier
had produced a convincing argument \cite{scale} that the FF approach
incorporates AF and we could not see how a discrete
model, freezing at nonzero temperature, could possibly be
AF, we decided to refine our data by concentrating
on the region $p/m<13$. Our results \cite{pl2} showed small but 
statistically significant deviations between $O(3)$ and FF, but
excellent agreement between $O(3)$ and the dodecahedron.

The comparison of the FF and $O(3)$ could have been marred by
lattice artefacts, a problem to which we will return below. This is why
a different comparison was performed by Patrascioiu \cite{quasi},
who computed the renormalized spin 2-point function versus the
physical distance $x/\xi$. The results were very similar with
the ones we will report here about the icosahedron. They suggested
that the dodecahedron and $O(3)$ models share the same continuum limit.

Another reason to investigate the discrete spin models arose in 1999,
while in collaboration with Balog, Niedermaier, Niedermayer and Weisz
we decided to compare the FF prediction for the renormalized coupling $g_R$
(to be defined below) with its lattice continuum limit value. Although
our first results \cite{col1} suggested excellent agreement,
as we continued to reduce the error bars and especially to take data
at larger correlation length $\xi\approx 167$, our original extrapolation
to the continuum limit became dubious and in our second paper \cite{col2}
we stated that we could no longer give a reliable number for the
lattice continuum limit value of $g_R$. In the hope of gaining 
insight into this issue, we turned again to the discrete models. 
Corroborating our previous findings regarding $G_r(p/m)$ and 
$G_r(x/\xi)$, the data, to be shown later, suggested that $g_R$ also agreed 
in the dodecahedron, icosahedron and $O(3)$ models.

In the course of the debates of our collaboration F. Niedermayer raised
the intriguing possibility that maybe indeed all these models do have
the same continuum limit, which however is AF. The present paper is our 
reply to his suggestion. The results were communicated
to him already in 1999, hence we were surprised by his recent paper
with P. Hasenfratz \cite{hn}. In their paper, Hasenfratz and Niedermayer
claim that indeed they find strong numerical evidence that the dodecahedron
and icosahedron models have the same continuum limit as $O(3)$, but 
state that,
since {\it ``overwhelming evidence exists that the $O(3)$ model is
AF''}, the dodecahedron and icosahedron models must be AF too.
The authors do not mention which `overwhelming evidence' they have in
mind, but omit the following facts, which in our opinion are relevant
to this issue:
\begin{itemize}
\item The data for $g_R$ in $O(3)$ produced by Balog et al \cite{col2}  
at larger $\xi$ values {\it do not support} the $1/\xi$ fit for
the lattice artefacts.
\item The last report of Balog et al \cite{col2} retracted the original
prediction $g_R=6.77(2)$ and stated that due to uncontrollable
lattice artefacts, no continuum value could be given.
\item In our recent paper \cite{abs}, we combined mathematically rigorous 
arguments with some numerics to conclude that the $O(3)$ model
must undergo a transition to a massless phase at finite $\beta$.
We then proved rigorously that such a phase transition rules out
the existence of AF in the massive continuum limit.
\end{itemize}
Being based on numerics, any of the above stated facts could be false, 
but if Hasenfratz and Niedermayer have evidence to that effect, they
should put it forth so that it can be scrutinized.

In this paper we will compare the continuum limit of the $O(3)$
model to that of the icosahedron model by comparing the values of
$g_R$ and of the renormalized spin 2-point function $G_r(x/\xi)$. 
The advantage of the icosahedron model is the existence of a rather
well localisable critical point, whereas, as stated above,
 the dodecahedral model
appears to have a soft intermediate phase. To address the issue
of AF, we study the value of the L\"uscher-Weisz-Wolff (LWW) coupling
constant at the critical point. According to LWW, AF requires
that the continuum limit of this observable vanishes at short distances.
If in fact
the icosahedron model has the same continuum limit as $O(3)$, then
the LWW coupling constant should vanish at the critical point
of the former model. We find excellent evidence that it does not.
Thus our conclusion is that either, in spite of the excellent 
agreement observed for $\xi\approx 121$,
 $O(3)$ and the icosahedron model have different 
continuum limits, or neither is AF.

The paper is organized as follows: we first describe the critical
properties of the icosahedon model and in particular locate its
critical point. This allows us to determine the value of the LWW
running coupling at the critical point. Then we move on to the comparison
of the icosahedron and the $O(3)$ models. We compare the renormalized
coupling constant and the renormalized spin-spin correlation function 
of the two models and present convincing evidence that they converge to 
the same continuum limit.

%22222222222222222222222222222222222222222222222222222222222222222222
\section{Critical Behavior of Icosahedron Model}

We first describe the
critical properties of the icosahedron model: we locate the critical
point and give some estimates of critical exponents. This allows us
to determine the value of the L\"uscher-Weisz-Wolff (LWW) coupling
constant at the critical point, which is independent ot the size $L$
of the lattice, as required by scaling. This (nonzero) value is also
the short distance limit of the continuum value of the LWW coupling
constant. 
 
The icosahedron model is defined by the standard nearest neighbor
coupling between the spins
\bee
H=-\sum_{\langle i j \rangle} s_i\cdot s_j
\ee
where the spins $s_i$ are unit vectors forming the vertices of
a regular icosahedron. In suitable coordinates, those 12 vertices
are given by
\bee
e_k=(s\cos{2\pi k\over 10},s\sin{2\pi k\over 10},c\cos(\pi k)) \ \
(k=1,2,...10);
\ee
where
\bee
s={2\over\sqrt{5}},\   c={1\over\sqrt{5}}
\ee
and
\bee
e_{11}=(0,0,1)\ \  {\rm and}\ \  e_{12}=(0,0,-1)
\ee

The model has at least two phases, a high temperature phase
with exponential clustering and full symmetry under the icosahedral
group $Y$, and a low temperature phase with spontaneous magnetization
and 12 coexisting phases, with the magnetization pointing into one
of the 12 directions of the icosahedron. At intermediate temperatures
there could be in principle a phase with partial breaking of $Y$, but
there is no reasonable candidate for a possible unbroken subgroup.
There is also the possibility of an extended intermediate phase with no
symmetry breaking, but actual enhancement of the symmetry from $Y$
to $O(3)$ as well as only algebraic decay of correlations analogous
to the $Z_N$ models and the dodecahedron model (see above).

In the icosahedron model the situation seems to be simpler, however: it
seems to have a single critical point separating the high and low 
temperature phases, as we will demonstrate.

To determine the critical point we proceed as follows: we determine the
mass gap $m(L)=1/\xi(L)$ in an (ideally infinite) strip of width $L$;
in practice we use a finite strip of size $L\times L_t$ with $L_t>>L$
and measure the effective correlation length defined by
\bee
\xi(L)={1\over 2\sin(\pi/L_t)}\sqrt{\chi/ G_1-1}
\ee
where 
\bee
\chi={1\over L L_t}\sum_{i,j}\langle s_i\cdot s_j\rangle,
\ee
\bee
G_1={1\over L L_t}\sum_{i,j}\langle s_i\cdot s_j\rangle
\exp[{2\pi (i_1-j_1)/L_t}].
\ee
We then study the behavior of the quantity
\bee
\bar g\equiv {2L\over (N-1)\xi(L)},
\ee
which we consider as a function $\bar g(z,\xi)$ of $z=L/\xi(\infty)$ 
and the infinite volume correlation length $\xi(\infty)$. 
The continuum limit of this quantity for fixed $z$ is
the `running coupling constant' introduced by L\"uscher, Weisz and Wolff
\cite{LWW} for the $O(N)$ models. In the high temperature phase, where 
the model has a mass gap $m(\infty)=1/\xi(\infty)$ in the infinite 
volume limit, $\bar g$ will grow linearly with $L$. On the other hand
in the low temperature magnetized phase, the mass gap in a finite
volume goes to 0 faster than  $1/L$, so $\bar g$ will decrease to
$0$ as $L\to\infty$. At a critical point the model will be scale
invariant for large distances and $\bar g$ should converge to a finite
nonzero limit.

\begin{figure}[htb]
\centerline{\epsfxsize=8.0cm\epsfbox{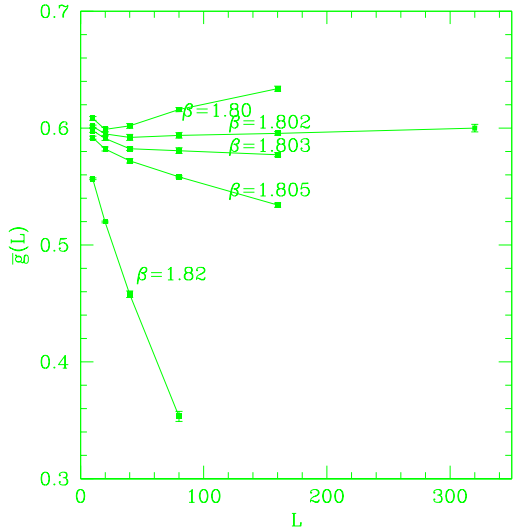}}
\caption{\it The LWW coupling constant $\bar g$ as a function of $L$ 
for various	values of $\beta$}
\label{glcr}
\end{figure}

To determine $\bar g$ we took data on lattices of  size $L\times L_t$ 
with $L_t=10 L$; $L$ was varied from 10 to 320.  
Fig.\ref{glcr} (Table 1)
shows clearly the dramatic change in behavior between $\beta=1.800$ and 
$\beta=1.805$. For $\beta=1.802$ $\bar g$ shows only some small
variation for small $L$ ($L<40$) and stabilizes for larger $L$.
So we estimate
\bee
\beta_{crt}=1.802(1)
\ee
where the error is of course somewhat subjective.

\begin{figure}[htb]
\centerline{\epsfxsize=8.0cm\epsfbox{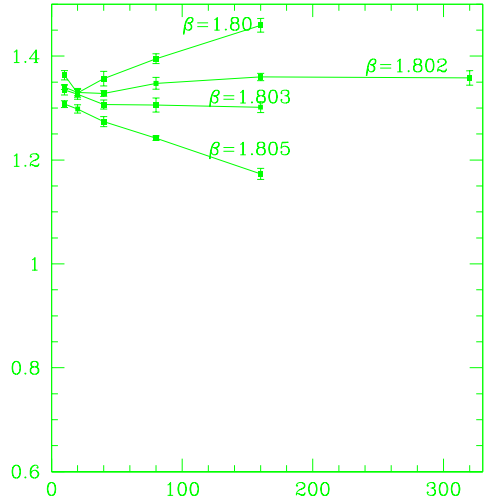}}
\caption{\it The renormalized coupling for asymmetric $L\times L_t$
lattices as a function of $L$ for various values of $\beta$}
\label{gras}
\end{figure}

To corroborate this determination of the critical point, we also 
measured on the same lattices the renormalized coupling defined as
\bee
g_R=\left({5\over 3}-{g_4\over g_2^2}\right){LL_t\over\xi^2}.
\ee
Here $g_2$ is the magnetic susceptibility multiplied by the volume of
the lattice $L\times L_t$,
i.e.
\bee
g_2=\sum_{i,j}\langle s_i\cdot s_j \rangle
\ee
and
\bee
g_4=\sum_{i_1,i_2,i_3,i_4} \langle (s_{i_1} \cdot s_{i_2})
(s_{i_3} \cdot s_{i_4})\rangle
\ee
This quantity is also a Renormalization Group (RG) invariant and
therefore should also go to a constant for $\beta=\beta_{crt}$.
It is well known that in the high temperature phase $g_R(L)$ goes to a 
nontrivial constant for $L\to\infty$, (which as about 6.7, see below)
whereas in the magnetized phase
it goes to $0$, so one expects a qualitatively similar picture as for 
$\bar g$. The data presented in Table 1 and displayed in Fig.\ref{gras}
confirm this nicely and are consistent with the estimate of 
$\beta_{crt}$ given above.

Finally we want to see if our determination of $\beta_{crt}$ is
consistent with a singularity in the thermodynamic values
of the correlation length $\xi$ and the susceptibility $\chi$.
We therefore measured $\xi$ and $\chi$ on lattices with $L/\xi\approx 7$
for various values of $\beta<1.802$; our data are given in Tab.2 .
There is a row listing the number of runs; a run consists of 100,000
cluster updates for thermalization followed by 20,000 sweeps of the
lattice for measurements. Each run is started independently with a
randomly chosen new configuration.

\begin{figure}[htb]
\centerline{\epsfxsize=8.0cm\epsfbox{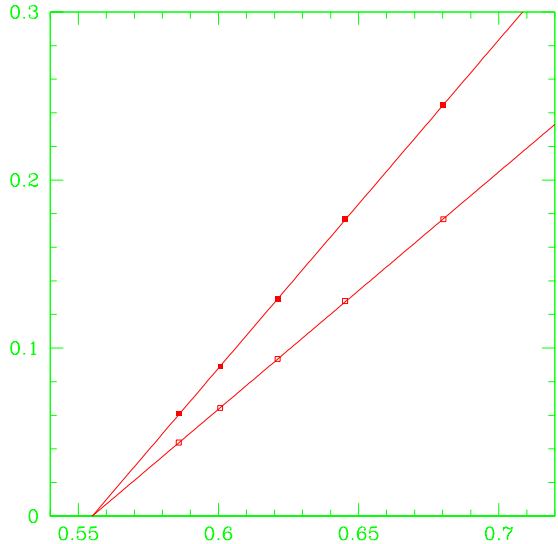}}
\caption{\it Thermodynamic values of $\xi^{-1/\nu}$ and $\chi^{-1/\gamma}$
with $\nu=1.717$ and $\gamma=3.002$ as functions of $\beta$; the lines
are fits.}
\label{critt}
\end{figure}

To describe the critical behavior of the data for $\xi$ and $\chi$
two types of fits were tried: firstly a Kosterlitz type fit with an 
exponential singularity 
\bee
\xi=C_\xi \exp\left(-a_\xi\over\sqrt{\beta_{crt}-\beta}\right),\ \
\chi=C_\chi \exp\left(-a_\chi\over\sqrt{\beta_{crt}-\beta}\right),
\ee
and
secondly a power law fit of the type
\bee
\xi=C_\xi(\beta_{crt}-\beta)^{-\nu}, \ \
\chi=C_\chi(\beta_{crt}-\beta)^{-\gamma}
\ee
or similar ones in $1/\beta$. Both types of fit are not very good,
with similar quite large values of $\chi^2$ and they do not allow a very
precise determination of the fit parameters. The reason seems to be
the following: the asymptotic singular behavior seems to have significant
subleading contributions, which however cannot be well determined with
only 5 values of $\beta$. Trying to fit our very precise data with
functions that do not describe the behavior with similar accuracy
necessarily leads to a poor fit quality, even though visually the
data may be very well described (see Fig.\ref{critt}).

The Kosterlitz type fit leads to predictions of 
$\beta_{crt}\approx 1.93$, which is unacceptably large -- this value is
deeply in the magnetized phase. The power law fits, on the other hand, 
give values of $\beta_{crt}$ quite close to our preferred value 1.802 .
By playing with the number of parameters and fitting in $1/\beta$ as
well as $\beta$ we obtain quite spread in the values of the exponents;
they fall into the intervals
\bee
 \nu=1.6 - 2.0 \ \ {\rm and}\ \  \gamma=2.9 - 3.5
\ee
This is consistent with a value of 
\bee
\eta=2-{\gamma\over\nu}\approx .25
\ee
which is also favored by the data for spin-spin correlation function
(see below).

In Fig.\ref{critt} we use the best values produced by the power law
fit in $1/\beta$ with no subleading corrections, namely
\bee
\nu=1.717,\ \ \gamma=3.002.
\ee
We plot $\xi^{-1/\nu}$ and $\chi^{-1/\gamma}$ vs $\beta$ together
with the fits, which are straight lines intersecting the abscissa at 
$\beta=1.802$.
So even though the thermodynamic data for $\xi$ and $\chi$ do not
lead to a precise prediction of the critical point and the critical
exponents, they are certainly consistent with our determination
based on the LWW coupling constant. 

We also investigated the possibility that the transition from the
high temperature phase to the one with long range order is first
order, but we did not find any signal for phase coexistence.

%33333333333333333333333333333333333333333333333333333333333333
\section{The renormalized coupling in the icosahedron and $O(3)$ models}

To check whether the icosahedron and the $O(3)$ model define the same 
continuum limit, we also determined the renormalized coupling constant
$g_R$ on thermodynamic lattices. For this purpose we took
data on square lattices with $z=L/\xi\approx 7$. To correct
for the small finite size effects still present, we determined a
finite size scaling curve by taking data at various values of $z=L/\xi$
at $\beta=1.550$ corresponding to $\xi\approx 19.7$. The finite
size effects are fitted with the function
\bee
g_R(z)=g_R(\infty)(1-d \sqrt{z}\exp(-z).
\ee

\begin{figure}[htb]
\centerline{\epsfxsize=8.0cm\epsfbox{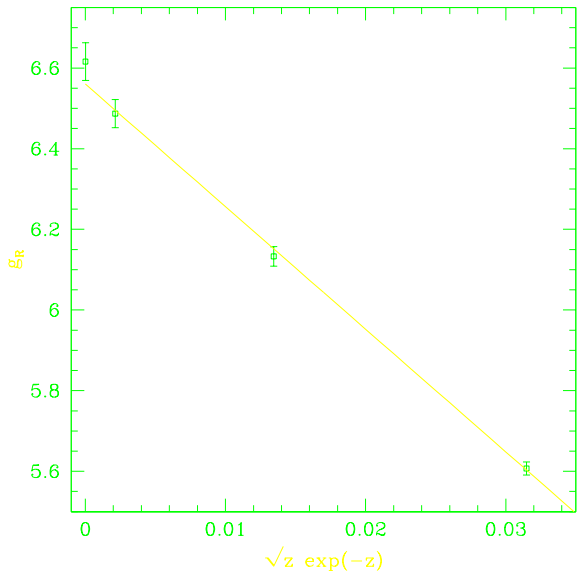}}
\caption{\it Finite size scaling of $g_R$}
\label{fss}
\end{figure}

This behavior is inspired by the spherical model and was successfully
used already in \cite{col1,col2}. In Fig.\ref{fss} we show the
data at $\beta=1.55$ together with the fit, which has a $\chi^2$ per
d.o.f. of 1.1 ; the constant $d$ is determined to be
\bee
d=4.63\pm .15
\ee
The results of this fit are then used to extrapolate our data taken
with $z\approx 7$ to the thermodynamic values. In Tab.2 we present our
data together with the extrapolation.

\begin{figure}[htb]
\centerline{\epsfxsize=8.0cm\epsfbox{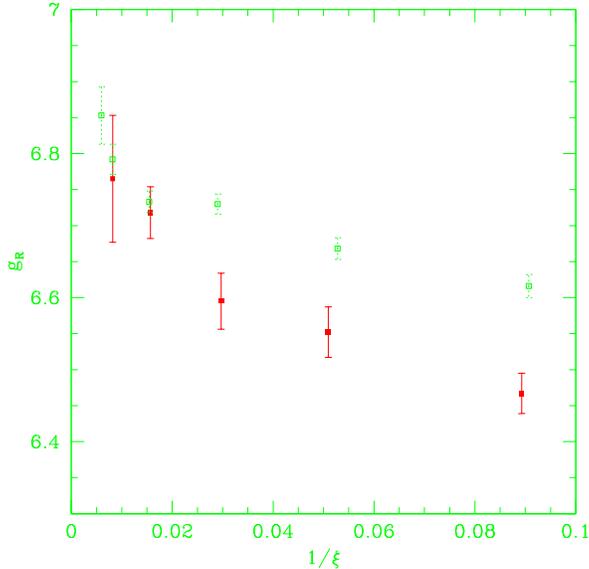}}
\caption{\it Comparison of $g_R$ in the icosahedron and $O(3)$ models
Full symbols:icosahedron; open symbols: $O(3)$}
\label{comp}
\end{figure}

Finally we compare our data for $g_R$ in the icosahedron model with 
those we had obtained earlier for the $O(3)$ model \cite{col2}
and again presented in Tab.2. Fig. \ref{comp} shows the thermodynamically
extrapolated  data for the two models for various values of $\xi$.
Even though the lattice artefacts are
quite different for the two models, and in spite of the fact that we
are not quite sure how one should extrapolate to the continuum limit,
the data show that the two models approach each other with
increasing $\xi$ (decreasing lattice spacing) and suggest that
they will have the same continuum limit.

%44444444444444444444444444444444444444444444444444444444444444

\section{Spin correlation function in the icosahedron and $O(3)$ models}

In this section we compare the renormalized spin-spin correlation
functions of the icosahedron and the $O(3)$ models. They are defined
as
\bee
G(i/\xi)\equiv {\xi^2\over \chi}\langle s(0)\cdot s(i)\rangle
\ee

\begin{figure}[htb]
\centerline{\epsfxsize=8.0cm\epsfbox{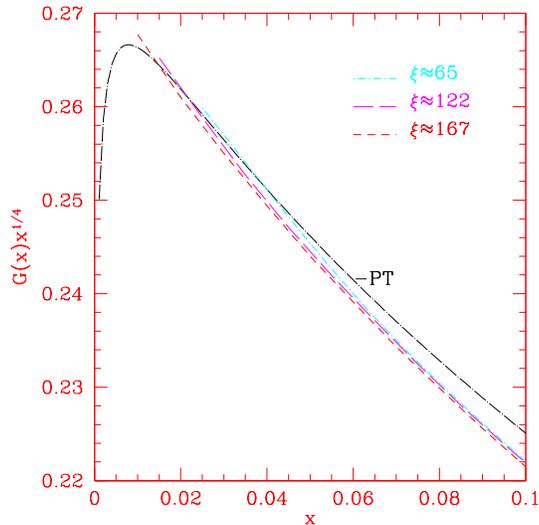}}
\caption{\it The renormalized spin-spin correlation for the $O(3)$
model for various lattice spacings}
\label{gxo3}
\end{figure}

\begin{figure}[htb]
\centerline{\epsfxsize=8.0cm\epsfbox{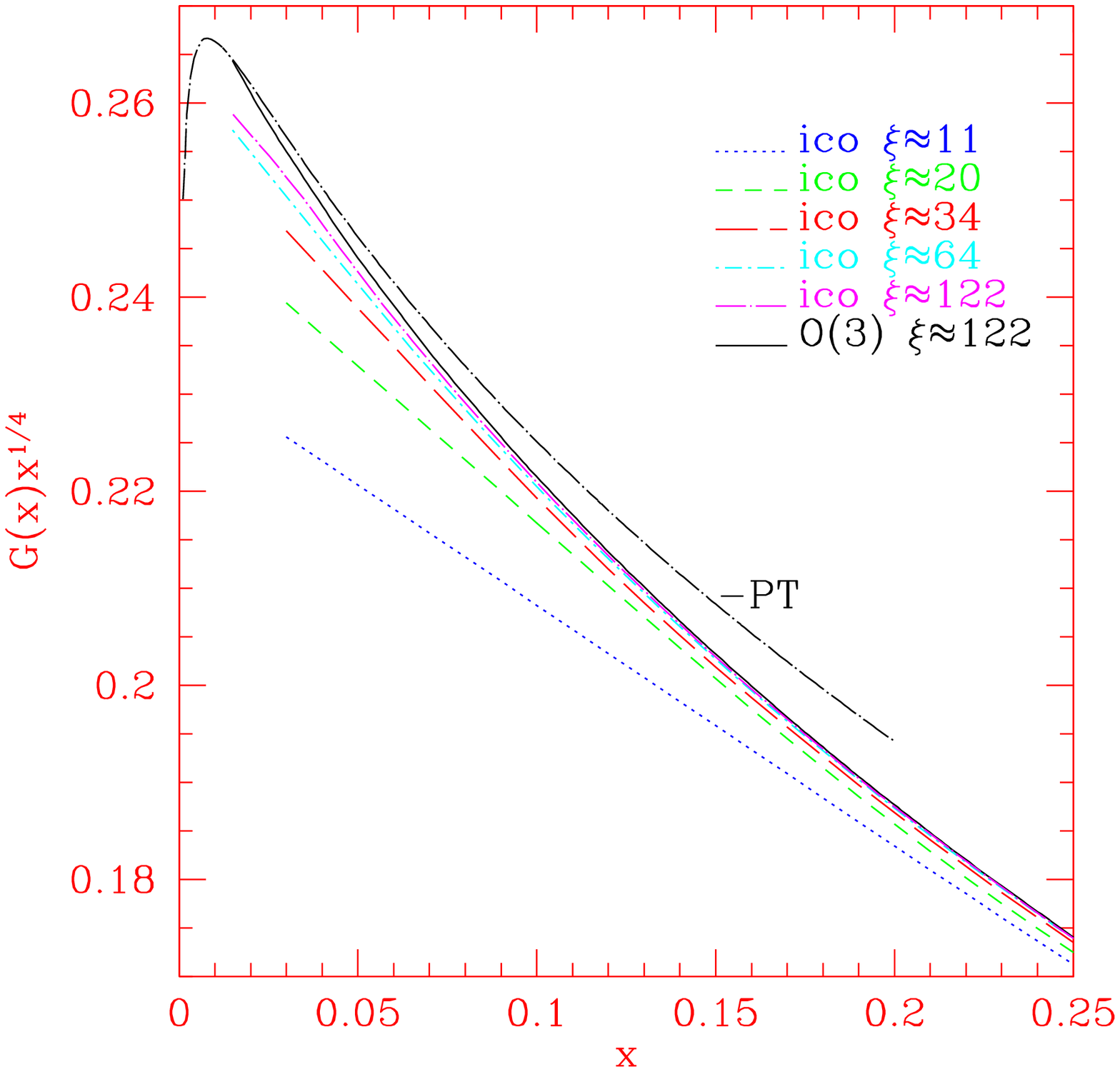}}
\caption{\it The renormalized spin-spin correlation for the icosahedron
model for various lattice spacings: solid line $O(3)$ model for 
$\xi\approx 122$}
\label{gxico}
\end{figure}
\noindent
The physical distance is
\bee
x={i\over\xi}
\ee
We measured the two-point functions in the icosoahedron model
at $\beta=1.47$, 1.55, 1.61, 1.665 and 1.707 corresponding to 
$\xi\approx$ 11, 20, 34, 64, 122 (see Tab.2). For the $O(3)$ model
we used the data from ref.\cite{quasi} at $\beta=1.8$, 1.9, 1.95 
corresponding to
$\xi\approx 65$, 122 and 168. In Fig.\ref{gxo3} we show
$G(x)x^{1/4}$ for the $O(3)$ model together with the 2-loop PT prediction
\cite{bn}

\bee
G(x)={1.000034657\over 3\pi^3 1.001687^2}\left[t+\ln t+1.1159+{1\over t}
\ln t+{.1159\over t}\right]
\ee
where $t=-\ln (x/8)-1$.
It can be seen that the data approach their continuum limit from above
and deviate considerably from the PT prediction. For $x>0.4$, barring
some very slow convergence to the continuum limit, our data suggest
that the lattice artefacts are quite small for the large correlation
lengths we are using.

In Fig.\ref{gxico} we present $G(x)x^{1/4}$ for the icosahedron model
together with the same expression for $O(3)$ at approximately
the largest value of $\xi\approx 122$. The lattice artefacts have
the opposite sign at least for the lattices with $\xi<122$, i.e.
the data are increasing with decreasing lattice spacing. At 
$\xi\approx 122$ they are already quite close to the corresponding
correlation function in the $O(3)$ model. It is therefore reasonable
to expect that if we could further refine the lattice we would see that 
the data follow the behavior of the $O(3)$ model, i.e. the lattice 
artefacts would change sign and the final approach to the continuum 
limit would be from above.

It should be stressed that already at $\xi\approx 122$ the icosahedron
and $O(3)$ data are much closer to each other than the $O(3)$ data are
to the PT prediction.

%55555555555555555555555555555555555555555555555555555555555555
\section{Conclusion: Universality between the icosahedron and 
$O(3)$ models and asymptotic freedom}

We have accumulated strong evidence that the continuum limits
of the discrete icosahedron model and the continuous classical
Heisenberg ($O(3)$) model describe the same quantum field theory.

\begin{figure}[htb]
\centerline{\epsfxsize=8.0cm\epsfbox{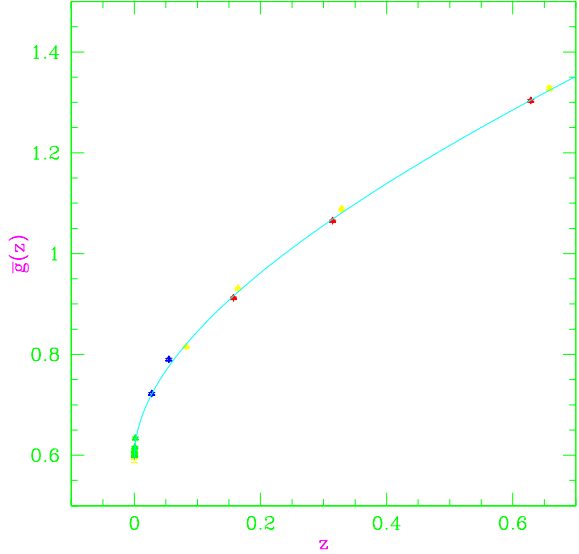}}
\caption{\it The LWW running coupling for the iscosahedron model
as a function of $L/\xi(\infty)$}
\label{run}
\end{figure}

As discussed in the introduction, this is one more fact which puts
the asymptotic freedom of the $O(3)$ model severely into doubt.
The point of view advocated by Hasenfratz and Niedermayer \cite{hn},
namely that the continuum limit of the discrete icosahedron model
should be asymptotically free is untenable in view of our results
about the LWW running coupling $\bar g$: our data (see Fig.\ref{glcr})
indicate that $\bar g(L)$ runs to a fixed point value $g^\ast\approx .59$
at small distances. Actually to determine the true running coupling,
one should take first the continuum limit at fixed $z=L/\xi$ and then 
the limit
$z\to 0$. Since this is not feasible, we instead studied the finite
size scaling at and around the critical point and took as our estimate
of $g^\ast$ the apparent limit
\bee
\lim_{L\to\infty} m(L)L
\ee
at the critical point. Somebody might wonder
if it is not possible that the continuum limit shows asymptotic
freedom after all; in other words, if it is possible that
\bee
\lim_{z\to 0} \lim_{\xi\to\infty} \bar g(z,\xi)=0 \ \ ?
\ee
Our data make this extremely unlikely: for fixed
$\beta<\beta_{crt}\approx 1.802$ $\bar g$ is {\it increasing} with
$L$ (except for very small lattices) and 
it is always larger than .59, even at $\beta=1.802$, our estimated
critical point. To claim that in the continuum limit $\bar g(z,\infty)$ would go to 0 
one would have to assume some truly bizarre $L$ dependence at fixed 
$\beta$. This is illustrated in our Fig.\ref{run} which shows some of 
our data for $\bar g$ as a function of $z=L/\xi(\infty)$. We used
data at $\beta=1.665$ and 1.707, where we know the correlation length
$\xi(\infty)$ quite well, together with the data taken slightly below
the estimated critical point, wehre we used the fit appearing in
Fig.\ref{critt} to estimate $\xi(\infty)$. The solid curve is a fit
of the form
\bee
\bar g(z)=g^\ast +a z^{1/2}+bz
\ee

Since we did not make any effort to control the lattice artefacts
and estimate the precise continuum values, this figure should 
be taken with some caution. It does, however, illustrate nicely
the qualitative behavior of the LWW running coupling near the
critical point.

To sum up: the universality observed between the icosahedron and
the $O(3)$ model gives strong evidence {\it against} asymptotic freedom 
of the latter.

\newpage
{\bf Tab.1:}\\
{\it The LWW running coupling $\bar g$ and the renormalized coupling
$g_R$ as a function of $L$ for various values of $\beta$}\\

$\beta=1.665$: \\
\begin{tabular}[t]{r||r|r|r}
$L$ & 10 & 20 & 40\\
\hline
$\bar g$ &.9121(14) & 1.0651(11) & 1.3033(21) \\
$g_R$ &2.216(13)&2.501(12) & 3.080(29) \\
\end{tabular}\\

$\beta=1.707$: \\
\begin{tabular}[t]{r||r|r|r|r}
$L$ & 10 & 20 & 40 & 80 \\
\hline
$\bar g$ &.8156(9) &.9312(12) & 1.0882(23) & 1.3276(35) \\
$g_R$ &1.898(6)&2.213(11) & 2.593(23) & 3.171(36)\\
\end{tabular}\\

$\beta=1.75$: \\
\begin{tabular}[t]{r||r|r}
$L$ & 10 & 20 \\
\hline
$\bar g$ &.7221(24)& .7902(21) \\
$g_R$ &1.695(14)&1.841(24)  \\
\end{tabular}\\

$\beta=1.80$: \\
\begin{tabular}[t]{r||r|r|r|r|r}
$L$ & 10 & 20 & 40 & 80 & 160 \\
\hline
$\bar g$ &.6083(16)& .5989(17) & .6018(21) & .6158(17) & .6337(22) \\
$g_R$ &1.363(9) &1.329(9)   &1.357(14)  & 1.395(9)  &1.459(13)  \\
\end{tabular}\\

$\beta=1.802$: \\
\begin{tabular}[t]{r||r|r|r|r|r|r}
$L$ & 10 & 20 & 40 & 80 & 160 & 320 \\
\hline
$\bar g$ &.6015(10)& .5951(14)& .5919(26)& .5938(24)& .5955(14)& .6000(32)\\
$g_R$ &1.340(4) &1.3330(6) &1.328(5)  &1.348(11) &1.360(7)  &1.358(14)\\
\end{tabular}\\

$\beta=1.803$:\\
\begin{tabular}[t]{r||r|r|r|r|r}
$L$ & 10 & 20 & 40 & 80 & 160 \\
\hline
$\bar g$ &.5979(16)& .5912(18)& .5823(14)& .5807(25)& .5772(17)\\
$g_R$ &1.335(10) &1.327(10) &1.307(9)  &1.306(14) &1.302(10)\\
\end{tabular}\\

$\beta=1.805$:\\
\begin{tabular}[t]{r||r|r|r|r|r}
$L$ & 10 & 20 & 40 & 80 & 160 \\
\hline
$\bar g$ &.5918(13)& .5821(17)& .5721(19)& .5583(8)& .5341(19)\\
$g_R$ &1.308(7)  &1.298(8)  &1.274(10) &1.242(4)  &1.173(11)\\
\end{tabular}\\

$\beta=1.82$:\\
\begin{tabular}[t]{r||r|r|r|r}
$L$ & 10 & 20 & 40 & 80 \\
\hline
$\bar g$ &.5566(8)& .5196(8)& .4578(28)& .3533(42)\\
$g_R$ &1.213(4)&1.119(4) &.964(10)  &.652(8) \\
\end{tabular}\\

\newpage
{\bf Tab.2:}\\
{\it Correlation length $\xi$, susceptibility $\chi$ and renormalized
coupling $g_R$ in the high temperature phase of the icosahedron model}\\
\begin{tabular}[t]{r||r|r|r|r|r}
$\beta$ & 1.470 & 1.550 & 1.610 & 1.665 & 1.707 \\
\hline
\hline
$\L$   &      80   &        140 &       250  &      500   & 910 \\
\hline
$\xi$  & 11.203(5) & 19.627(12) & 33.655(21) & 63.628(33) & 122.09(16)\\
\hline
$\chi$ &181.88(10) &479.43(36)  & 1228.04(91)&3774.5(2.1) &12009(17)\\
\hline
$g_R(L)$ &6.404(28)&6.487(35)   & 6.546(39)  & 6.684(36)  & 6.715(88)\\
\hline
$g_R(\infty)$&6.467(28)&6.552(35) & 6.595(39)  & 6.718(36)  & 6.765(88)\\
\hline
\# runs & 200      &        100 &       100  &      149   &  28\\
\end{tabular}\\

{\bf Tab.3:}\\
{\it Correlation length $\xi$ and renormalized coupling $g_R$ in the 
high temperature phase of the $(O3)$ model (from \cite{col2}}\\
\begin{tabular}[t]{r||r|r|r|r|r|r}
$\beta$ & 1.5   & 1.6   & 1.7   & 1.8   & 1.9  & 1.95 \\
\hline
\hline
$L$   &      80   &        140 &       250  &      500   & 910 & 1230\\
\hline
$\xi$  & 11.030(7) & 18.950(14) & 34.500(15) & 64.790(26) & 122.330(74)
& 167.71(17)\\
\hline
$g_R(L)$ &6.553(16)&6.612(15)&6.665(14)&6.691(15)&6.737(21)&6.792(40)\\
\hline
$g_R(\infty)$&6.616(16)&6.668(15)&6.730(14)&6.733(15)&6.792(21)&6.853(40)\\
\hline
\# runs & 344   &   370 &   367  &   382  & 127 & 68\\
\end{tabular}\\

%\widetext

\begin{thebibliography}{99}
%
\bibitem{LWW}
M. L\"uscher, P. Weisz and U. Wolff, {\sl Nucl. Phys.} {\bf B 359} (1991)
221.
%
\bibitem{jose}
V. Jos\'e, L. Kadanoff, S. Kirkpatrick and D. Nelson, {\sl Phys. Rev. B} 
{\bf 16} (1977) 1217.
%
\bibitem{fs}
J. Fr\"ohlich and T. Spencer, {\sl Commun. Math. Phys.}
{\bf 83} (1982) 411.
%
\bibitem{ns}
C. Newman and L. Schulman, {\sl Phys. Rev. B} {\bf 26} (1982) 3910.
%
\bibitem{pat}
A. Patrascioiu, {\sl Phys. Rev. Lett.} {\bf 54} (1985) 2292.
%
\bibitem{prs}
A.Patrascioiu, J. Richard  and E.Seiler, {\sl Phys. Lett.} {\bf B241}
(1990) 229.
%
\bibitem{dod}
A.Patrascioiu, J. Richard  and E.Seiler, {\sl Phys. Lett.} {\bf B254}
(1991) 173.
\bibitem{pl1}
A. Patrascioiu and E. Seiler, {\sl Phys. Lett.} {\bf B445} (1998) 160.
%
\bibitem{bn} 
J.Balog and M.Niedermaier, 
{\sl Nucl. Phys.} {\bf B500} (1997)421.
% 
\bibitem{scale} J.Balog and M.Niedermaier,
{\sl Phys.Rev.Lett.} {\bf 78} (1997) 4151.
%
\bibitem{pl2}
A.Patrascioiu and E.Seiler, {\sl Phys. Lett.} {\bf B430} (1998) 314.
%
\bibitem{quasi}
A.Patrascioiu, {\it Quasi-asymptotic freedom in the teo dimensional
$O(3)$ model}, AZPH-TH-99-01, hep-lat/0002012 
%
\bibitem{col1} 
J. Balog, M. Niedermaier, F. Niedermayer, A.Patrascioiu, E. Seiler and
P. Weisz, {\sl Phys. Rev.} {\bf D60} (1999) 094508.
A.Patrascioiu, 
%
\bibitem{col2} 
J. Balog, M. Niedermaier, F. Niedermayer, A.Patrascioiu, E. Seiler and
P. Weisz, {\sl Nucl.Phys.} {\bf B583} (2000) 614.
%
\bibitem{hn}
P. Hasenfratz and F. Niedermayer, {\it Unexpected results in 
asymptotically free quantum field theories},
BUTP-2000-15, hep-lat/0006021.
%
\bibitem{abs}
A.Patrascioiu and E.Seiler, {\it Absence of asymptotic freedom in 
nonabelian models}, MPI-PHT-2000-07, hep-th/0002153.
%
\end{thebibliography}
\end{document}